\documentclass{ws-mpla}
\usepackage{cite}
\usepackage{graphicx}
\usepackage{amsmath}
\usepackage{url}

\newcommand{\aspi}{\frac{\hat\alpha_s}{\pi}}

\begin{document}

\markboth{J. ERLER, P. MASJUAN, H. SPIESBERGER}
{CHARM MASS FROM A PAIR OF SUM RULES}

\catchline{}{}{}{}{}

\title{CHARM QUARK MASS DETERMINED FROM A PAIR OF SUM RULES\\
}

\author{\footnotesize JENS ERLER}

\address{Instituto de F\'isica, Universidad Nacional Aut\'onoma de M\'exico\\
Apartado Postal 20--364, M\'exico D.F. 01000, M\'exico\\
erler.jens@gmail.com}

\author{PERE MASJUAN\footnote{Speaker.}}

\address{ PRISMA Cluster of Excellence, Institut f\"ur Kernphysik,\\
 Johannes Gutenberg-Universit\"at, 55099 Mainz, Germany\\
 masjuan@kph.uni-mainz.de
}

\author{HUBERT SPIESBERGER}

\address{ PRISMA Cluster of Excellence, Institut f\"ur Physik,\\
  Johannes Gutenberg-Universit\"at, 55099 Mainz, Germany,\\
    and Centre for Theoretical and Mathematical Physics and 
  Department of Physics,\\
    University of Cape Town, Rondebosch 7700, South Africa\\
    spiesber@uni-mainz.de
    }

\maketitle

\pub{Published 4 November 2016}{}

\begin{abstract}
In this paper, we present preliminary results of the determination of the
 charm quark mass $\hat{m}_c$ from QCD sum rules of 
moments of the vector current correlator calculated in perturbative QCD at 
${\cal O} (\hat \alpha_s^3)$. Self-consistency between two different sum rules allow to
determine the continuum contribution to the moments without requiring experimental input,
except for the charm resonances below the continuum threshold. 
The existing experimental data from the continuum region is used, then, to confront the 
theoretical determination and reassess the theoretic uncertainty.

\keywords{QCD Sum rules; Charm mass}
\end{abstract}

\ccode{PACS Nos.: 12.38.-t, 12.38.Lg, 14.65.Dw}

\section{Introduction}

In this paper, we propose to revisit the method of relativistic sum rules
to extract the charm quark mass with
emphasis on the evaluation of the uncertainty. 

Among the most precise charm mass determinations, including
lattice simulations~\cite{Laschka:2011zr,Namekawa:2011wt,Heitger:2013oaa,Carrasco:2014cwa,Chakraborty:2014aca} and deep-inelastic scattering data~\cite{Alekhin:2013nda,Abramowicz:1900rp}, relativistic QCD sum rules play an important role on establishing the
quark mass at the few-percent level~\cite{Novikov:1977dq,Shifman:1978bx, 
Shifman:1978by,Chetyrkin:2010ic,Dehnadi:2011gc,Bodenstein:2011ma,Narison:2011rn}. The method, based on rigorous field 
theoretical principles, can be systematically improved.
Nevertheless, the resulting uncertainties are dominated by theory 
errors which are notoriously difficult to estimate and often subject 
of vigorous debate. 

In this talk, based on the work of Ref.~\cite{Inprep}, we would stress that  the overall error may also be constrained 
within our approach to the QCD sum rules. To this end, we will adopt a strategy where 
the only exploited experimental information are the masses and 
electronic decay widths of the narrow resonances in the sub-continuum 
charm region, $J/\Psi(1S)$ and $\Psi(2S)$. 

Consistency between two different QCD sum rules will be seen to suffice to constrain the 
continuum of charm pair production with good precision. For this 
procedure to work it is crucial to include alongside the first or 
second moment sum rules also the zeroth moment, as the latter 
exhibits enhanced sensitivity to the continuum. 

Comparison with existing data on the $R$-ratio for hadronic relative to leptonic 
final states in $e^+ e^-$ annihilation will then serve as a control, 
providing an independent error estimate which we interpret as the 
error on the method and (conservatively) add it as an 
{\em additional\/} error contribution. In this way, we can show 
that the overall precision in $\hat{m}_c$ from relativistic sum 
rules is at the sub-percent level.

Further details and discussions about the method and results presented in this work can be found in Ref.~\cite{Inprep}.

\section{Defining the zeroth sum rule}

Let us consider the transverse part of the correlator $\Pi_q(t)$ 
of two heavy-quark vector currents. $\Pi_q(t)$ obeys the subtracted 
dispersion relation

\begin{equation}
12\pi^2 ( \Pi_q (0) -  \Pi_q (-t) )= t  \int_{s_0}^\infty \frac{{\rm d} s}{s}   \frac{R_q(s)}{s +t } \ ,
\label{SR}
\end{equation}
\noindent
where we have defined $12\pi$Im$[\Pi_q (t + i \epsilon)] = R_q(t)$. By the optical theorem, $R_q(s)$ can be related to the measurable cross section for heavy-quark production in $e^+e^-$ annihilation. The lower limit of the integral $s_0$ is fixed from the threshold for heavy quark production which is the unknown quantity we want to determine. Ultimately is the function $R_q(s)$ which will decide what exact value to use for $s_0$ since
\begin{equation}\label{ResCont}
R_q(s) =
  \begin{cases}
    R_q^{\textrm{Res}}(s)  & \text{if $0< s_0 < 4 M^2$}\\
    R_q^{\textrm{Cont}}(s) & \text{if $4 M^2 \le s_0 < \infty$}
 \end{cases}
\end{equation}
\noindent
where $R_q^{\textrm{Res}}(s)$ contains a finite set of narrow resonances produced below the heavy-flavor production threshold, and $R_q^{\textrm{Cont}}(s)$ describes the continuum production above that threshold. As soon as the resonance contribution $R_q^{\textrm{Res}}(s) $ is separated from $R_q(s)$, $s_0$ is identified with the open charm threshold $s_0 = 4 M^2$ with $M=M_{D^0}=1864.84$MeV.\\

Assuming now global quark-hadron duality, we can write~\cite{Inprep}:
\begin{equation}
 \int_{s_0}^\infty \frac{{\rm d} s}{s(s+t)} R_q(s) 
=  \int_{s_0}^\infty \frac{{\rm d} s}{s(s+t)} R_q^{\textrm{pQCD}}(s) 
\label{qHD}
\end{equation}
where $R_q^{\textrm{pQCD}}(s)$ corresponds to the $R_q(s)$ ratio calculated in perturbative QCD (pQCD) order by order in the $\alpha_s(s)$ expansion. Eq.~(\ref{qHD}) together with Eq.~(\ref{SR}), implies:
\begin{equation} 
 \Pi_q (0) -  \Pi_q (-t) 
= 
 \hat{\Pi}_q^{\textrm{pQCD}} (0) - \hat{ \Pi}^{\textrm{pQCD}}_q (-t) 
\end{equation}
where $\hat{\Pi}^{\textrm{pQCD}}_q (t)$ is the correlator $\Pi_q(t)$ calculated in pQCD and the caret indicates the $\overline{\textrm{MS}}$ scheme. $\hat{\Pi}^{\textrm{pQCD}}_q (t)$  obeys a subtracted dispersion relation, then, given by 
\begin{equation}
12\pi^2 \frac{  \hat{\Pi}_q^{\textrm{pQCD}} (0) - \hat{ \Pi}^{\textrm{pQCD}}_q (-t)  } {t}=   \int_{4 \hat{m}_q^2}^\infty \frac{{\rm d} s}{s}   \frac{R_q(s)}{s +t } \ ,
\label{SRpQCD}
\end{equation}
\noindent
where $\hat{m}_q = \hat{m}_q(\hat{m}_q)$ is the mass of the heavy quark. Equation~(\ref{SRpQCD}) defines a set of sum rules that allow us to define theoretical ${\cal M}_n^{\textrm{th}}$ and experimental ${\cal M}_n^{\textrm{exp}}$ moments~~\cite{Novikov:1977dq,Shifman:1978bx, 
Shifman:1978by}:
\begin{equation}\label{SRM}
{\cal M}_n^{\textrm{th}} 
= \frac{12 \pi^2}{n!}\frac{{\rm d}^n }{{\rm d} t^n} \hat \Pi_q (t) \bigg|_{t=0} 
= \int_{s_0}^\infty {\rm d} s  \frac{R_q(s)}{s^{n+1} } 
= {\cal M}_n^{\textrm{exp}}\, .
\end{equation}

We can also define the zeroth moment~\cite{Erler:2002bu,Inprep} ${\cal M}_0^{\textrm{th,exp}}$ by taking the limit $\lim_{t \to \infty}$ in Eq.~(\ref{SRpQCD}):
\begin{equation}
{\cal M}_0^{\textrm{th}} \, = \,12\pi^2 \lim_{t \to \infty} \frac{  \hat{\Pi}_q^{\textrm{pQCD}} (0) - \hat{ \Pi}^{\textrm{pQCD}}_q (-t)  } {t}
=\lim_{t \to \infty} \int_{4 \hat{m}_q^2}^\infty \frac{{\rm d} s}{s}\frac{ R_q(s)}{s+t}= \, {\cal M}_0^{\textrm{exp}}\\
\label{SRlim}
\end{equation}

After taking the $\lim_{t \to \infty}$ and multiplying by $t$, Eq.~(\ref{SRlim}) as it stands is not well defined neither for ${\cal M}_0^{\textrm{th}}$ nor for ${\cal M}_0^{\textrm{exp}}$ for which they must be regularized. At a given order in pQCD, the required regularization can be obtained 
by subtracting the zero-mass limit of $R_{q}(s)$, which we write as 
$3 Q_q^2 \lambda_1^q(s)$ with $Q_q$ the quark charge. $\lambda_1^q(s)$ 
is known up to ${\cal O}(\hat \alpha_s^4)$, but we will only need the 
third-order expression~\cite{Chetyrkin:2000zk},
\begin{eqnarray} \nonumber
  \lambda^q_1(s) 
  &=& 1 
  + \frac{\hat \alpha_s}{\pi}
  + \frac{\hat \alpha_s^2}{\pi^2} \left[ \frac{365}{24} - 11 \zeta(3)
     + n_q \left( \frac{2}{3} \zeta(3) - \frac{11}{12} \right)  \right] 
  \\ 
  &+& \frac{\hat \alpha_s^3}{\pi^3} 
     \left[\frac{87029}{288} - \frac{121}{8} \zeta(2) 
     - \frac{1103}{4} \zeta(3) + \frac{275}{6}\zeta(5) 
     \right. 
  \\  \nonumber
  &+&  \left. n_q \left(- \frac{7847}{216} + \frac{11}{6} \zeta(2) 
  + \frac{262}{9} \zeta(3) - \frac{25}{9}\zeta(5) \right) 
  + n_q^2 \left(\frac{151}{162} - \frac{\zeta(2)}{18} 
  - \frac{19}{27} \zeta(3) \right)  \right],
\label{lambda1}
\end{eqnarray}
where $\hat \alpha_s = \hat \alpha_s (s)$, $n_q = n_l + 1$ 
and $n_l$ is the number of light flavors (taken as massless), 
{\em i.e.}, quarks with masses below the heavy quark under 
consideration. 

Let us then define the function $H(t)$ to have exactly the same large $t$ behavior as $ \hat{\Pi}_q (t)$ (including leading divergence terms such as $\log(-t/\mu^2)$) up to ${\cal O}(\hat \alpha_s^3)$. This is, explicitely, $\lim_{t\to \infty} ( \hat{\Pi}^{\textrm{pQCD}}_q (t)  - H(t) ) =0 + {\cal O}(\hat \alpha_s^4)$. By definition, Im$[H(s+i \epsilon)] = \frac{3Q_q^2}{12\pi}\lambda_1^q(s)$ as the leading divergent terms from $ \hat{\Pi}_q (t)$ come from the massless limit of $R_q(s)$.  \\

\noindent
$H(t)$ satisfies a subtracted dispersion relation:
\begin{equation}\label{Hdis}
H(0)  -  H(-t) = \frac{t}{\pi}  \int_{\mu^2}^\infty \frac{{\rm d} s}{s}   \frac{\textrm{Im}[H(s)]}{s + t} 
\end{equation}
where, as we have said, Im$[H(s+i \epsilon)] = \frac{3Q_q^2}{12\pi}\lambda_1^q(s)$. The lower limit of the integral is such that after integrating over Im$[H(s+i \epsilon)]$, the leading logarithms from the large $t$ behavior of $ \hat{\Pi}_q (t)$, i.e., the $\log(-t/\mu^2)$'s, are exactly recovered.

With all these definitions, we can cancel the divergences in Eq.~(\ref{SRlim}) by subtracting Eq.~(\ref{Hdis}) from it~\cite{Inprep}:
\begin{eqnarray} \label{RegM0}
12 \pi^2 & \lim_{t\to \infty} & \frac{ \hat  \Pi^{\textrm{pQCD}}_q (0) - \hat \Pi^{\textrm{pQCD}}_q (-t) - (H(0)  - H(-t))}{t} \\ \nonumber
&=& \lim_{t\to \infty}  \int_{4 \hat m^2}^\infty \frac{{\rm d} s}{s}   \frac{R_q(s) - 12 \pi^2 \textrm{Im}[H(s)]}{s+t}  
-   \int_{\mu^2}^{4 \hat m^2} \frac{{\rm d} s}{s} \frac{ 12 \pi^2 \textrm{Im}[H(s)] }{s+t}\, .\\ \nonumber
\end{eqnarray}

Equation~(\ref{RegM0}) defines the regularized zeroth moment. As we have said, the optical theorem relates $R_{q}(s)$ in Eq.~(\ref{ResCont}) with the cross section for heavy-quark production in $e^+e^-$ annihilation. Below the threshold for continuum heavy-flavor production, 
$R_q^{\rm Res}(s)$ is approximated by $\delta$-functions \cite{Novikov:1977dq},
\begin{equation}
  R_q^{\rm Res}(s) = 
  \frac{9\pi}{\alpha_{\rm em}^2(M_R)} M_R \Gamma_R^e\delta(s-M_R^2) .
\label{Rres}
\end{equation}
The masses $M_R$ and electronic widths $\Gamma^e_R$ of the resonances 
\cite{Agashe:2014kda} are listed in Table~\ref{ResPDG} and 
$\alpha_{\rm em}(M_R)$ is the running fine structure constant 
at the resonance\footnote{The values for $\alpha_{\rm em}(M_R)$ 
were determined with help of the program {\tt hadr5n12} \cite{jegerlehner-url}.}.
To parametrize $R_q^{\rm Cont}(s)$,  we assume that continuum 
production can be described on average by the simple ansatz 
\cite{Erler:2002bu,Inprep}
\begin{equation}
  R_q^{\rm Cont}(s)
  = 3 Q^2_q \lambda^q_1 (s) 
  \sqrt{1 - \frac{4\, \hat{m}_q^2 (2 M)}{s^\prime}} 
  \left[ 1 + \lambda^q_3 \frac{2\, \hat{m}_q^2(2 M)}{s^\prime} 
  \right]\, ,
\label{ansatz}
\end{equation}
where $s' := s + 4(\hat{m}_q^2(2M) - M^2)$, and $M$ is taken as the 
mass of the lightest pseudoscalar heavy meson, {\em i.e.}, $M = 
M_{D^0} = 1864.84$ MeV for charm quarks~\cite{Agashe:2014kda}. 
$\lambda^q_3$ is a constant to be determined. Equation~(\ref{ansatz}) 
interpolates smoothly between the threshold and the onset of open 
heavy-quark pair production and coincides asymptotically with the 
prediction of pQCD for massless quarks. 

Performing the limit $\lim_{t\to \infty}$ in Eq.~(\ref{RegM0}) will allow us to define the zeroth sum rule. Doing so, we need the results of Refs.~\cite{Chetyrkin:1996cf} and \cite{Chetyrkin:1997un} 
and $R_q(s) = \sum\limits_{\rm resonances} R_q^{\rm Res}(s) + 
R_q^{\rm cont}(s)$. Multiplying then by $t/3Q_q^2$, and setting $\mu^2=\hat m^2$, the zeroth sum rule reads~\cite{Inprep}:
\begin{eqnarray} \label{SR0} \nonumber
 &&  \sum\limits_{\rm resonances} 
  \frac{9\pi\Gamma^e_R}{3 Q_q^2 M_R \hat\alpha_{em}^2 (M_R)} 
  + \int\limits_{4 M^2}^\infty \frac{{\rm d} s}{s}  \left(
   \frac{R_q^{\rm Cont}(s)}{3 Q_q^2} -\lambda_1^q(s)\right)
  - \int\limits_{\hat{m}_q^2}^{4M^2}
   \frac{{\rm d} s}{s} \lambda^q_1 (s) 
  =\\ 
  = &&- \frac{5}{3} + \aspi \left[ 4 \zeta(3) - \frac{7}{2} \right] 
  \\    \nonumber
 & +& \left(\aspi \right)^2  
   \left[ \frac{2429}{48} \zeta(3) - \frac{25}{3} \zeta(5) 
   - \frac{2543}{48} 
   + n_q \left( \frac{677}{216} - \frac{19}{9} \zeta(3) \right) 
   \right] + \left(\aspi \right)^3 A_3 ,
\end{eqnarray}
where $\hat\alpha_s = \hat \alpha_s(\hat m_q)$. The third-order 
coefficient $A_3$ is available in numerical form~\cite{Kiyo:2009gb, 
Hoang:2008qy},
\begin{equation}
A_3 = -9.863 + 0.399 \, n_q - 0.010 \, n_q^2\ .
\end{equation}
Notice that the continuum $R_q^{\rm Cont}(s)$ contributes with the 
lower integration limit $4M^2$, while the subtraction term 
$\lambda_1^q(s)$ is integrated starting from $\hat{m}_q^2$. 

To perform the integral $\int\limits_{4 M^2}^\infty \frac{{\rm d} s}{s}  \left(\frac{R_q^{\rm Cont}}{3 Q_q^2} -\lambda_1^q(s)\right)$ in Eq.~(\ref{SR0}), it is convenient to expand first $\lambda_1^q(s)$ in $\hat{\alpha}_s$ using the RGE for $\hat \alpha_s(s)$ selecting the reference scale as $\hat m_q^2$. In this way, the integral over $s$ becomes trivial and well defined, i.e., with all the divergences in both $R_q^{\rm Cont}(s)$ and $\lambda_1^q(s)$ are removed. 

Equation~(\ref{SR0}) contains two unknowns, the quark mass $\hat m_q(\hat m_q)$ and the parameter $\lambda_3^q$ entering in our prescription for $R_q^{\rm Cont}(s)$. The zeroth sum rule is the most sensitive to the continuum region and shall be used to determine $\lambda_3^q$. Self-consistency with another moment sum rule can then be used to determine the quark mass. 
 
\begin{table}[t]
\tbl{
Resonance data~\cite{Agashe:2014kda}
used in the analysis. 
The uncertainties from the resonance masses are negligible 
for our purpose.
}
{\begin{tabular}{@{}crr@{}} \toprule
$R$ & $M_R$ [GeV] & $\Gamma_R^e$ [keV] \\
\hline
 \rule[-2mm]{0mm}{7mm}
 $J/\Psi(1S)$ & 3.096916 & 5.55(14)  \\
 \rule[-2mm]{0mm}{6mm}
 $\Psi(2S)$ & 3.686109 & 2.36(4)  \\
\hline
\end{tabular}}
\label{ResPDG}
\end{table}

Theory predictions for the higher moments in perturbative QCD can 
be cast into the form 
\begin{equation}
  {\cal M}_n^{\rm pQCD}  = 
  \frac{9}{4} Q_q^2  
  \left( \frac{1}{2\hat{m}_q(\hat{m}_q)} \right)^{2n}  \hat{C}_n 
\label{Mth}
\end{equation}
with
\begin{equation}
  \hat{C}_n = 
  C_n^{(0)} 
  + \left(\aspi \right)C_n^{(1)} 
  + \left(\aspi \right)^2C_n^{(2)} 
  + \left(\aspi \right)^3C_n^{(3)} 
  + {\cal O}(\hat \alpha_s^4) \, .
\label{Cnth}
\end{equation}
The $C_n^{(i)}$ are known up to ${\cal O}(\hat \alpha_s^3)$ for $n \leq 3$ 
\cite{Chetyrkin:2006xg,Boughezal:2006px,Maier:2008he,Maier:2009fz}, 
and up to ${\cal O}(\hat \alpha_s^2)$ for the rest~\cite{Chetyrkin:1997mb, 
Maier:2007yn}.  Since we need all the moments up to ${\cal O}(\hat \alpha_s^3)$ we use 
the predictions for $n>3$ provided in Ref.~\cite{Greynat:2011zp}. 
Once ${\cal M}_n^{\rm pQCD}$ are available, the determination of both $\hat m_q(\hat m_q)$ and $\lambda_3^q$
come from solving the system of the two equations. In this last step, the theoretical moments
are equated with their corresponding experimental counterparts defined in Eq.~(\ref{SRM}), i.e., 
${\cal M}_n^{\rm pQCD} = {\cal M}_n^{\rm exp}$ for $n>0$.
 
In general, vacuum expectation values of higher-dimensional operators 
in the operator product expansion (OPE) contribute to the moments of the 
current correlator as well. These condensates may be important for 
a high-precision determination of heavy-quark masses, in particular 
in the case of the charm quark. The leading term involves the 
dimension-4 gluon condensate 
\cite{Novikov:1977dq}, 
\begin{equation}
  {\cal M}^{\rm cond}_n 
  = 
  \frac{12 \pi^2 Q_c^2}{(4 \hat m_c^2)^{n+2}} 
  \langle \frac{\hat \alpha_s}{\pi} G^2 \rangle \, 
  a_n \left(1 + \frac{\hat \alpha_s(\hat m_c^2)}{\pi} b_n\right) \, .
\end{equation}
The coefficients $a_n$ and $b_n$ can be found in Refs.~\cite{Chetyrkin:2010ic,Broadhurst:1994qj}  
and \cite{Kuhn:2007vp}. In our fits we use the central value 
$\langle \frac{\hat \alpha_s}{\pi} G^2 \rangle^{\rm exp} 
= 0.014$~GeV$^4$ with an uncertainty of 
$\Delta \langle \frac{\hat \alpha_s}{\pi} G^2 \rangle = 0.014$~GeV$^4$, 
taken from the recent analysis \cite{Dominguez:2014fua}. 

\begin{figure}[h!]
\begin{center}
\includegraphics[width=\textwidth]{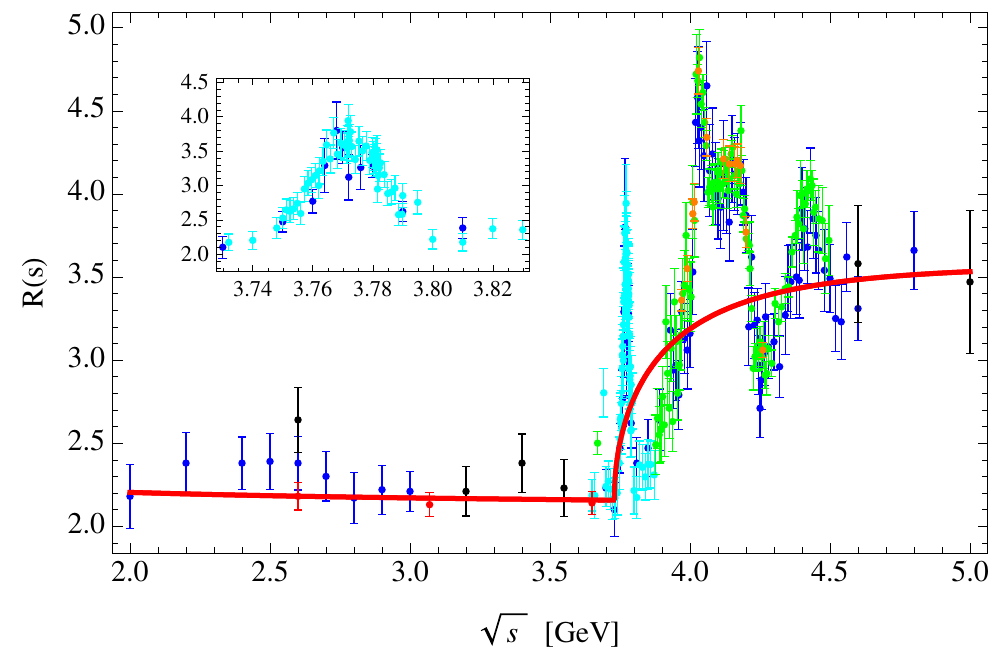}
\caption{
Data for the ratio $R$ for $e^+e^- \rightarrow $ hadrons in the 
charm threshold region: Crystal Ball CB86 (green) 
\cite{Osterheld:1986hw}; BES00, 02, 06, 09 (black, blue, 
 cyan,  and red) \cite{Bai:1999pk,Bai:2001ct,Ablikim:2006mb,Ablikim:2009ad}, and CLEO09 (orange) 
\cite{CroninHennessy:2008yi}. The full (red) curve shows 
$R_c^{\rm Cont}(s)$ with $\lambda_3^c =1.23$ and $\hat{m}_c(\hat{m}_c) 
= 1.272$GeV. The inner plot is a zoom-in into the energy range $2 M_{D^0} 
\leq \sqrt{s} \leq 3.83$ GeV.
}
\label{Rc}
\end{center}
\end{figure}

Including the condensate contribution when equating Eqs.~(\ref{SRM}) and (\ref{Mth}), i.e.,  $ {\cal M}_n^{\rm exp} = {\cal M}_n^{\rm pQCD} $  together with the zeroth sum rule, we determine values for the heavy quark mass 
$\hat{m}_c(\hat{m}_c)$ and the constant $\lambda^c_3$. The other 
moments are then fixed and can be used to check the consistency of 
our approach. No experimental data other than the resonance parameters 
in Table~\ref{ResPDG} are necessary. From the combination $0{\rm th}+2{\rm nd}$ sum rules, we obtain $\lambda_3^c =1.23$ and $\hat{m}_c(\hat{m}_c) = 1.272$GeV without errors as they come from solving a system of two equations. The error estimation is discussed in the next section.
Once both $\hat{m}_c(\hat{m}_c)$ and $\lambda^c_3$ are determined, 
we can compare our prescription for $R_q^{\rm Cont}(s)$ with experimental data in the threshold region, Fig.~\ref{Rc}. The full red curve
shows $R_c^{\rm Cont}(s)$ with $\lambda_3^c =1.23$ and $\hat{m}_c(\hat{m}_c) 
= 1.272$GeV and should be understood as an average determination of the cross-section in the threshold region.

\subsection{Uncertainty estimate}

In order to determine an error for the continuum contributions we proceed in the following way~\cite{Inprep}: instead of using 
Eqs.\ (\ref{SR0}, \ref{Mth}), we can compare 
experimental data shown in Fig.~\ref{Rc} with the zeroth moment in the restricted energy 
range of the threshold region, $2 M_{D^0} \leq \sqrt{s} \leq 4.8$~GeV 
to obtain an {\em experimental} value for $\lambda_3^c$, denoted 
$\lambda_3^{c,{\rm exp}}$. Here we fix $\hat{m}_c(\hat{m}_c)$ 
using Eq.~(\ref{Mth}) and proceed to solve Eq.~(\ref{SR0}) by comparing with ${\cal M}_0^{\rm exp}$. Then we can also determine an error, 
$\Delta \lambda_3^{c,{\rm exp}}$ from the experimental uncertainty 
of the data in this threshold region.

\begin{figure}[h]
\begin{center}
\includegraphics[trim=5cm 5cm 5cm 5cm, clip=true, width=0.9\textwidth]{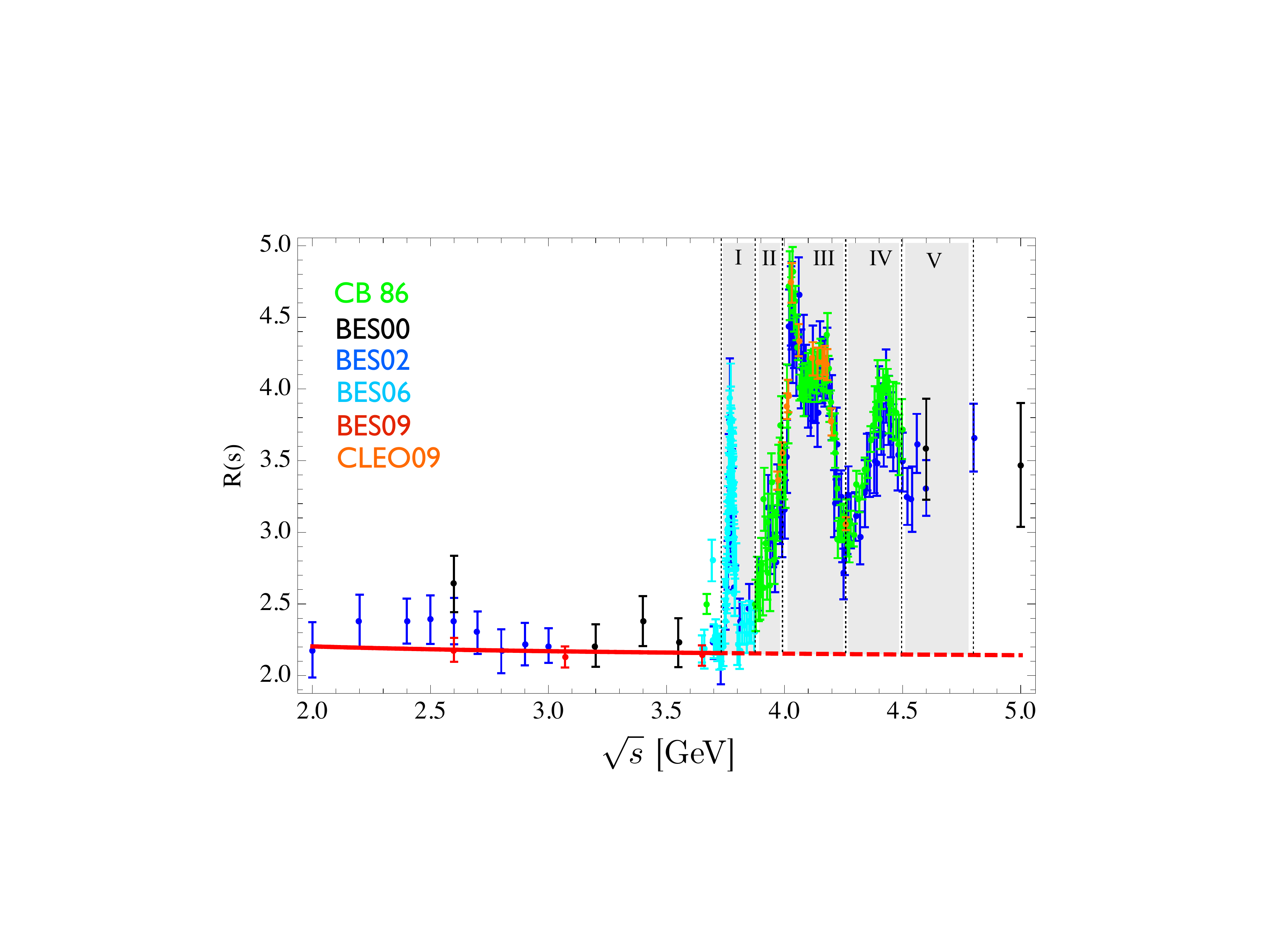}
\caption{
Data for the ratio $R$ for $e^+e^- \rightarrow $ hadrons in the 
charm threshold region: Crystal Ball CB86 (green) 
\cite{Osterheld:1986hw}; BES00, 02, 06, 09 (black, blue, 
 cyan,  and red) \cite{Bai:1999pk,Bai:2001ct,Ablikim:2006mb,Ablikim:2009ad}, and CLEO09 (orange) 
\cite{CroninHennessy:2008yi}. The gray bands indicate the five intervals considered for evaluating
the experimental moments.
}
\label{Fig:Intervals}
\end{center}
\end{figure}

We calculate the experimental moments via numerical integrals over
the available experimental data, cf. Fig.~\ref{Rc}. Experimental data is classified in five different intervals, see Fig.~\ref{Fig:Intervals},
which allow us to fully take into account correlated and uncorrelated uncertainties among different
collaborations and intervals. The results for the experimental moments in the threshold region, $2 M_{D^0} \leq \sqrt{s} \leq 4.8$~GeV 
are given in Table~\ref{table:Data1}. For the results in the columns labeled 'Data', light-quark contributions have 
been subtracted using the pQCD prediction at order ${\cal O} (\hat \alpha_s^3)$, see Ref.~\cite{Inprep}. Its second column shows the required value for $\lambda_3^{c,{\rm exp}}$ such that the zeroth moment sum rule is experimentally  satisfied after fixing $\hat{m}_c(\hat{m}_c) 
= 1.272$GeV. The rest of the moments in this column are reported to show the consistency of the approach. Even for the highest moments, the consistency is very good. The last column collects, for comparison, the value for the moments in the same energy region using $\lambda_3^c=1.23$ extracted from the theoretical determination.

\begin{figure}[h]
\begin{center}
\includegraphics[trim=7cm 7cm 7cm 7cm, clip=true, width=0.9\textwidth]{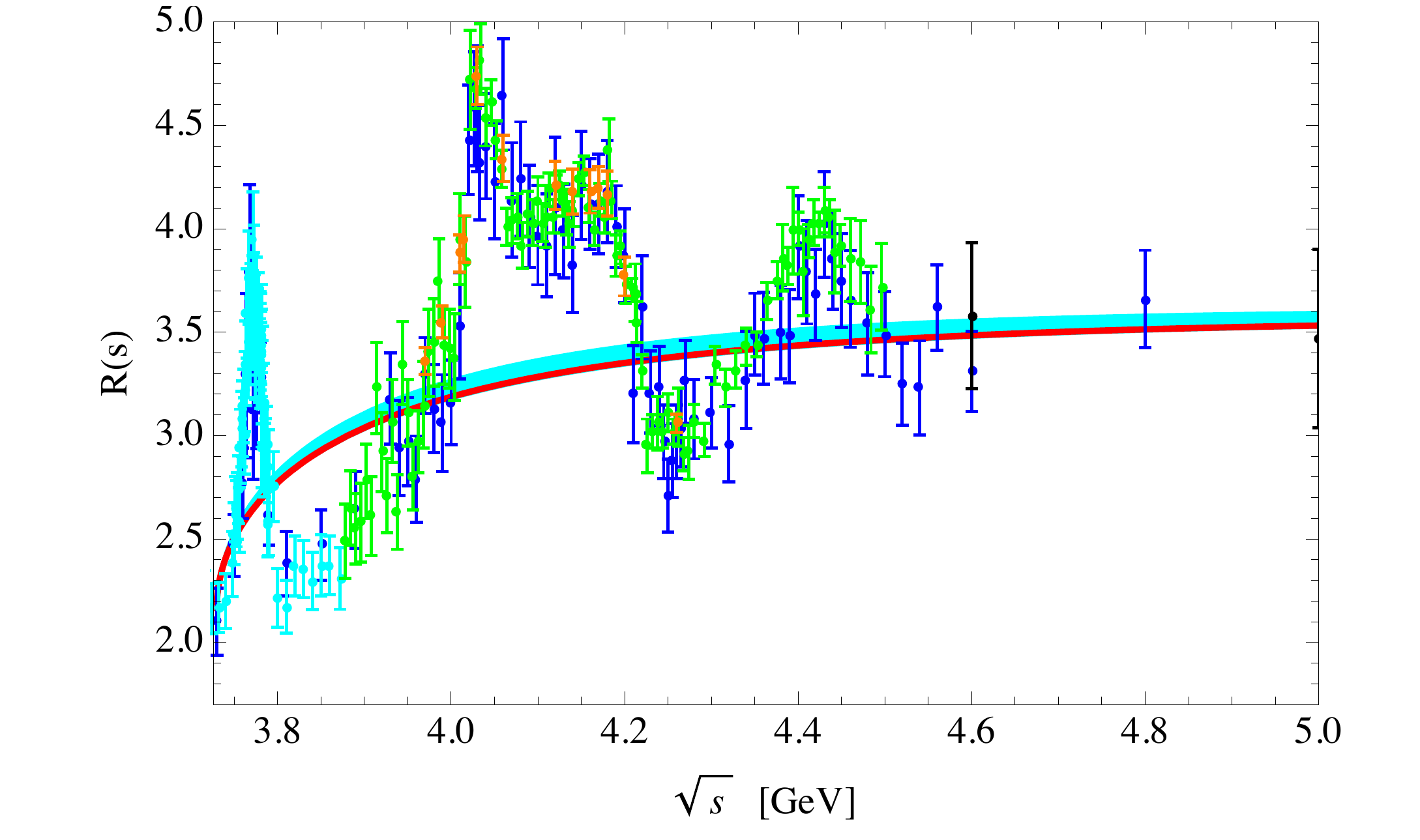}
\caption{
Data for the ratio $R$ for $e^+e^- \rightarrow $ hadrons in the 
charm threshold region: Crystal Ball CB86 (green) 
\cite{Osterheld:1986hw}; BES00, 02, 06, 09 (black, blue, 
 cyan,  and red) \cite{Bai:1999pk,Bai:2001ct,Ablikim:2006mb,Ablikim:2009ad}, and CLEO09 (orange) 
\cite{CroninHennessy:2008yi}. The full (red) curve shows 
$R_c^{\rm Cont}(s)$ with $\lambda_3^c =1.23$ and $\hat{m}_c(\hat{m}_c) 
= 1.272$GeV and the cyan band shows $R_c^{\rm Cont}(s)$ with $\hat{m}_c(\hat{m}_c) 
= 1.272$GeV and  $\lambda_3^{c,{\rm exp}} =1.34(17)$ .
}
\label{Rcshift}
\end{center}
\end{figure}

The shift in the moments resulting from the different values for $\lambda_3^{c}$ (either 
from two moments combined with resonance data only, or from the 
comparison of the 0th moment with continuum data in the threshold 
region) turns out to be small. Strictly speaking this shift is a 
one-sided error, but to be conservative we include it as an 
additional error in the results of Table \ref{Tab:MomentsBudget}. A graphical account of this shift is
shown in Fig.~\ref{Rcshift} as a cyan band for the result of the $0{\rm th}+2{\rm nd}$ moments pair for $\hat{m}_c(\hat{m}_c) 
= 1.272$GeV. In this case, $\lambda_3^{c,{\rm exp}} = 1.34(17)$, c.f. Table~\ref{Tab:MomentsBudget}.
 The red solid curve corresponds to the same pair of moments and the same quark mass with $\lambda_3^{c} = 1.23$, and well overlaps with the cyan band.

\begin{table}[t]
\tbl{
Contributions to the charm moments ($\times 10^{n}\, \mbox{GeV}^{2n}$) 
from the energy range $2 M_{D^0} \leq \sqrt{s} \leq 4.8$~GeV. For the 
results in the columns labeled 'Data', light-quark contributions have 
been subtracted using the pQCD prediction at order ${\cal O} 
(\hat \alpha_s^3)$. The entries here are obtained from the separate 
contributions shown in Fig.~\ref{Fig:Intervals} taking into 
account the correlation of systematic errors within each experiment. 
The third column uses $\hat{m}_c = 1.272 {\rm GeV}$ 
and $\lambda_3^{c,{\rm exp}}$ determined by the zeroth experimental 
moment (see text for details). The last column shows the theoretical 
prediction for the moments using $\hat{m}_c = 1.272 {\rm GeV}$ and 
$\lambda^c_3 = 1.23$. 
}
{\begin{tabular}{@{}l c c c@{}} \toprule
$n$ & Data  & $\lambda^c_3 = 1.34(17)$ & $\lambda^c_3 = 1.23$ \\
\hline
\rule[-2mm]{0mm}{6mm}
0 & 0.6367(195) & 0.6367(195) & 0.6239 \\
\rule[-2mm]{0mm}{3mm}
1 & 0.3500(101)  & 0.3509(111)  & 0.3436 \\
\rule[-2mm]{0mm}{3mm}
2 & 0.1957(54)  & 0.1970(65)  & 0.1928 \\
\rule[-2mm]{0mm}{3mm}
3 & 0.1111(29)  & 0.1127(38)  & 0.1102 \\
\rule[-2mm]{0mm}{3mm}
4 & 0.0641(16)  & 0.0657(23)  & 0.0642 \\
\rule[-2mm]{0mm}{3mm}
5 & 0.0375(9)   & 0.0389(14)   & 0.0380 \\
\hline
\end{tabular}}
\label{table:Data1}
\end{table}

Finally, we assign a truncation error to the theory prediction of 
the moments following the method proposed in Ref.~\cite{Erler:2002bu} 
which considers the largest group theoretical factor in the next 
uncalculated perturbative order as a way to estimate errors,  
\begin{equation}
  \Delta {\cal M}_n^{(i)} 
  = \pm Q_q^2 
  N_C C_F C_A^{i-1} \, 
  \left(\frac{\hat\alpha_s (\hat{m}_q)}{\pi}\right)^i
  \left( \frac{1}{2 \hat{m}_q(\hat{m}_q)} \right)^{2n} 
\label{truncerror}
\end{equation}
($N_C = C_A = 3$, $C_F = 4/3$). At order ${\cal O}(\hat \alpha_s^4)$, 
this corresponds to an uncertainty of $\pm 48 (\hat \alpha_s/\pi)^4$ 
for $\hat{C}_n^{(4)}$ in Eq.\ (\ref{Mth}). 

For the moments with $n>3$ taken from Ref.~\cite{Greynat:2011zp} 
we have to include additional uncertainties specific to the 
method used to obtain predictions for ${\cal M}_n$. These errors 
are very small, but included for completeness.  

\begin{figure}[h!]
\begin{center}
\includegraphics[width=0.9\textwidth]{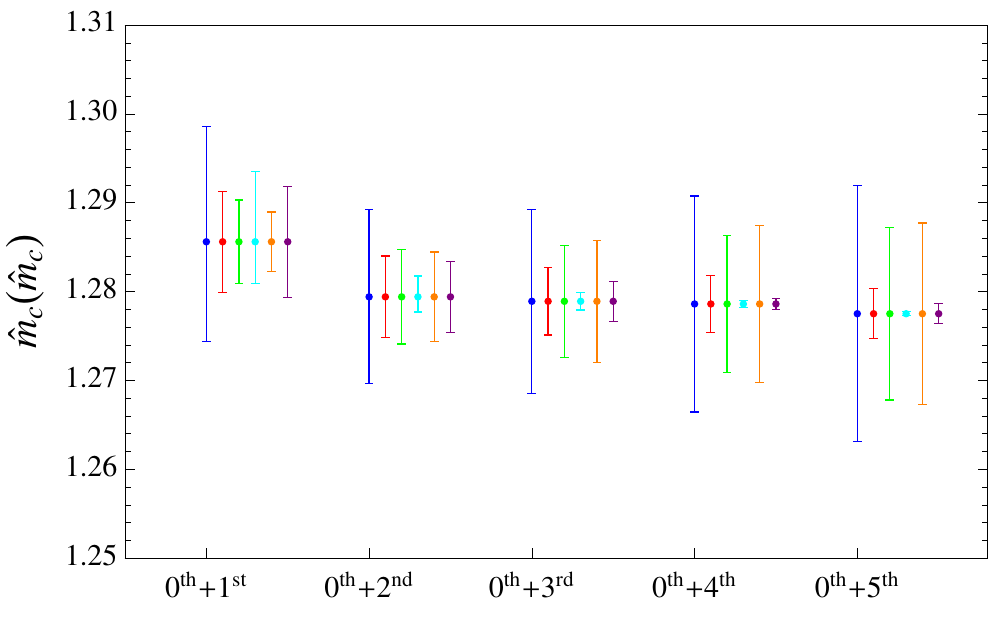}
\caption{
$\hat{m}_c(\hat{m}_c)$ using different combinations of moments 
and the error budget for each combination. Blue is the full error, 
red is the one from the resonance region, green from the truncation 
errors of the theoretical moments, cyan from the error of $\lambda_3^c$ 
(which is the combination of the shift and the experimental error on 
$\lambda_3^c$), orange from the gluon condensate and purple from the 
uncertainty of $\Delta \hat\alpha_s(M_z)$. 
}
\label{Fig:cmass}
\end{center}
\end{figure}

The charm mass and the continuum parameter $\lambda_3^c$ can, in 
principle, be determined from any combination of two moments, not only $0{\rm th}+2{\rm nd}$. The 
zeroth moment, however, is expected to provide the highest 
sensitivity. The results for combinations of the zeroth with one 
higher moment are summarized in Table \ref{Tab:MomentsBudget} and visualized in Fig.\ \ref{Fig:cmass}. 
We include the difference between the two possibilities to determine 
$\lambda_3^c$ as described above as an additional error. 

As an example on how to understand Table \ref{Tab:MomentsBudget}, select the $0{\rm th}+2{\rm nd}$ moments pair as the result for the quark mass, we would combine the total error $7.8$ MeV with $0.37 \times 14$ MeV error from the condensate uncertainty and with $2.6 \times 1.6$ MeV from $\hat \alpha_s(M_z)=0.1182(16)$~\cite{Agashe:2014kda}. Then, $\hat{m}_c(\hat{m}_c) = 1272(9)$MeV. Doing so for each pair of moments collected in the table, we notice that the combination $0{\rm th}+2{\rm nd}$ provides the smallest total error for the heavy quark mass.
Let us remark that for the highest moments, the truncation of the OPE series, i.e. condensates of higher dimension not considered in our approach can be important~\cite{Inprep}. While difficult to assert, we belief that these higher dimension condensates are well included in our condensate error estimate. However, to be on the safe side, charm quark mass determination using the fourth and fifth moment sum rules may have an underestimated error. They should not be considered. On the contrary, the zeroth and first moments are the ones less sensitive to the OPE truncation with the combination $0{\rm th} + 1{\rm st}$ being a most favorable choice. However, this combination is the one most sensitive to the continuum region, with largest shift in $\lambda_3^{c,{\rm exp}}$, cf. Table ~\ref{Tab:MomentsBudget}.  The pair of the zeroth and second moments is our optimal choice since balance well between reduced effects of the OPE series truncation and good description of the continuum region. This pair has also the smallest total uncertainty in the charm mass determination.

\begin{table}[h]
\tbl{Values of $\hat{m}_c(\hat{m}_c)$ and $\lambda_3^c$ determined 
from different pairs of moments and split-up of the errors for 
the charm mass. The errors denoted 'Total' are the quadratic 
sum of the two errors from $\lambda_3$, the one from the 
resonances and the truncation error. Errors induced by the 
uncertainties of $\hat \alpha_s$ and of $\langle \frac{\hat \alpha_s}{\pi} 
G^2 \rangle$ (in units of GeV$^4$) are parametric and are given separately in the 
last two lines. }
{\begin{tabular}{@{}lccccc@{}} \toprule
$\Delta \hat{m}_c(\hat{m}_c)~[\mbox{MeV}]$
& ${\cal M}_0$ -- ${\cal M}_1$ 
& ${\cal M}_0$ -- ${\cal M}_2$ 
& ${\cal M}_0$ -- ${\cal M}_3$ 
& ${\cal M}_0$ -- ${\cal M}_4$ 
& ${\cal M}_0$ -- ${\cal M}_5$ 
\\
\colrule
\rule[-2mm]{0mm}{7mm}
$\hat{m}_c(\hat{m}_c)~[\mbox{MeV}]$ 
& 1280.9 
& 1272.4 
& 1269.1 
& 1265.8 
& 1262.2 
\\ 
\rule[-2mm]{0mm}{7mm}
$\lambda_3^c$ 
& 1.154 
& 1.230 
& 1.262 
& 1.291 
& 1.323 
\\
\rule[-2mm]{0mm}{7mm}
$\lambda_3^{c,{\rm exp}}({\cal M}_0)$ 
& 1.35(17) 
& 1.34(17)
& 1.34(17) 
& 1.33(17)
& 1.32(17)
\\
\hline
\rule[-2mm]{0mm}{7mm}
Resonances 
& 5.8 
& 4.5 
& 3.9 
& 3.3 
& 2.8 
\\
\rule[-2mm]{0mm}{7mm}
Truncation error 
& 6.3
& 5.9
& 7.2
& 8.9
& 10.5
\\ 
\rule[-2mm]{0mm}{7mm}
Shift of $\lambda_3^c$ 
& +6.4 
& +1.5 
& +0.3 
& +0.1 
& +0.1 
\\
\rule[-2mm]{0mm}{7mm}
$\Delta \lambda_3^{c,{\rm exp}}$ 
& 4.7 
& 1.7 
& 0.7 
& 0.3 
& 0.2 
\\
\hline
\rule[-3mm]{0mm}{9mm}
Total:  
& 11.7 
& 7.8 
& 8.2 
& 9.5 
& 10.9 
\\
\hline
\rule[-2mm]{0mm}{7mm}
$10^3\times$Condensates 
& $-0.25 \langle \frac{\hat\alpha_s}{\pi} G^2 \rangle$ 
& $-0.37 \langle \frac{\hat\alpha_s}{\pi} G^2 \rangle$ 
& $-0.54 \langle \frac{\hat\alpha_s}{\pi} G^2 \rangle$ 
& $-0.73 \langle \frac{\hat\alpha_s}{\pi} G^2 \rangle$ 
& $-0.88 \langle \frac{\hat\alpha_s}{\pi} G^2 \rangle$ 
\\
& (-1.3 MeV)
& (-1.9 MeV)
& (-2.7 MeV)
& (-3.7 MeV)
& (-4.4 MeV)
\\
\hline
\rule[-2mm]{0mm}{7mm}
$10^3\times\Delta \hat\alpha_s(M_z)$ 
& $+3.6 \Delta \hat\alpha_s$ 
& $+2.6 \Delta \hat\alpha_s$ 
& $+1.6 \Delta \hat\alpha_s$ 
& $+0.6 \Delta \hat\alpha_s$ 
& $-0.4 \Delta \hat\alpha_s$ 
\\
& (+5.8 MeV)
& (+4.2 MeV)
& (+2.6 MeV)
& (+1.0 MeV)
& (-0.6 MeV)
\\
\rule[-2mm]{0mm}{3mm}
\botrule
\end{tabular}}
\label{Tab:MomentsBudget}
\end{table}

In Ref.~\cite{Erler:2002bu}, a determination of the heavy quark mass at ${\cal O}(\hat\alpha_s^2)$ was performed requiring as well self-consistency between the $0{\rm th}+2{\rm nd}$ moments to constrain the continuum region. The main difference between the results in Ref.~\cite{Erler:2002bu} and the ones presented here can be summarized as follows:  
\begin{itemize}
\item The theoretical sum rules are considered at order ${\cal O}(\hat\alpha_s^3)$, including the expression for $\lambda_3^q(s)$ and the theoretical moments.
\item The electronic widths of the narrow-width resonance states are measured now with higher precision. Actually, the shift in their central values amount to $1\sigma$ and induce the largest shift in the quark mass determination when comparing Table \ref{Tab:MomentsBudget} with Ref.~\cite{Erler:2002bu}.
\item The experimental determination of $\hat\alpha_s(M_z)$, i.e., the value $\hat\alpha_s(M_z)=0.1182(16)$ used in this work has been improved with respect to the one used in Ref.~\cite{Erler:2002bu}.
\item Other minor improvements include more experimental data in the continuum region and better determination of the condensate contribution.
\end{itemize}

\section{Conclusions}
\label{Sec:conclusions}
 
In this paper we presented a determination of the charm quark mass based on the work of Ref.~\cite{Inprep}. We revisit there the method of relativistic sum rules with emphasis on the evaluation of the uncertainty. By invoking the zeroth sum rule and requiring self-consistency with higher-moment sum rules, we can show that the overall error may be constrained within the approach. 

After considering the combination of two different sum rules, the only experimental information required are the masses and 
electronic decay widths of the narrow resonances in the sub-continuum 
charm region, $J/\Psi(1S)$ and $\Psi(2S)$. Comparison with experimental data in this region is later on used to check the results and determine an experimental error.

The results reported here are preliminary and more details are provided in Ref.~\cite{Inprep}.

\section*{Acknowledgments}

This work has been partially supported by DFG through the 
Collaborative Research Center ``The Low-Energy Frontier of 
the Standard Model" (SFB~1044). 
JE is supported by PAPIIT (DGAPA--UNAM) project IN106913 and 
by CONACyT (M\'exico) project 252167 and acknowledges financial 
support from the Mainz cluster of excellence PRISMA. 



\end{document}